\documentclass[
reprint,
 amsmath,amssymb,
 aps,
prd,
tightenlines
]{revtex4-1}
\usepackage[colorlinks,urlcolor=blue, citecolor=blue,linkcolor=blue]{hyperref}
\usepackage{siunitx}
\usepackage{booktabs}
\usepackage{color}
\usepackage{multirow}
\usepackage{overpic}
\usepackage{colortbl}
\usepackage{graphicx}
\usepackage{float}
\usepackage{array}
\usepackage{makecell}
\usepackage{subfigure}
\usepackage{graphicx}
\usepackage{dcolumn}
\usepackage{bm}
\usepackage{xcolor}
\usepackage{amsmath}

\definecolor{aa}{RGB}{0,0,139}

\newcommand{\chisq}{\chi^2}

\newcommand{\BR}{{\cal B}}

\newcommand{\etacp}{\eta_c(2S)}

\newcommand{\psip}{\psi(3686)}

\newcommand{\kk}{K^{+} K^{-}}

\usepackage{multirow}
\usepackage{enumerate}
\usepackage{amsmath}
\begin{document}


\title{\boldmath Observation of $\etacp \to \kk\eta$ }

\author{
M.~Ablikim$^{1}$, M.~N.~Achasov$^{4,c}$, P.~Adlarson$^{75}$, O.~Afedulidis$^{3}$, X.~C.~Ai$^{80}$, R.~Aliberti$^{35}$, A.~Amoroso$^{74A,74C}$, Q.~An$^{71,58,a}$, Y.~Bai$^{57}$, O.~Bakina$^{36}$
% ...
\\
\vspace{0.2cm}
(BESIII Collaboration)\\
\vspace{0.2cm} {\it
$^{1}$ Institute of High Energy Physics, Beijing 100049, People's Republic of China\\
% ...
$^{3}$ Bochum  Ruhr-University, D-44780 Bochum, Germany\\
$^{4}$ Budker Institute of Nuclear Physics SB RAS (BINP), Novosibirsk 630090, Russia\\
% ...
$^{35}$ Johannes Gutenberg University of Mainz, Johann-Joachim-Becher-Weg 45, D-55099 Mainz, Germany\\
$^{36}$ Joint Institute for Nuclear Research, 141980 Dubna, Moscow region, Russia\\
% ...
$^{57}$ Southeast University, Nanjing 211100, People's Republic of China\\
$^{58}$ State Key Laboratory of Particle Detection and Electronics, Beijing 100049, Hefei 230026, People's Republic of China\\
% ...
$^{71}$ University of Science and Technology of China, Hefei 230026, People's Republic of China\\
% ...
$^{74}$ University of Turin and INFN, (A)University of Turin, I-10125, Turin, Italy; (B)University of Eastern Piedmont, I-15121, Alessandria, Italy; (C)INFN, I-10125, Turin, Italy\\
$^{75}$ Uppsala University, Box 516, SE-75120 Uppsala, Sweden\\
% ...
$^{80}$ Zhengzhou University, Zhengzhou 450001, People's Republic of China\\
\vspace{0.2cm}
$^{a}$ Deceased\\
% ...
$^{c}$ Also at the Novosibirsk State University, Novosibirsk, 630090, Russia\\
% ...
}}
\date{\today}

\begin{abstract}
By analyzing $(27.12 \pm 0.14)\times10^{8}$ $\psip$ events accumulated with the BESIII detector, the decay $\eta_{c}(2S) \to K^{+} K^{-} \eta$ is observed for the first time with a significance of $6.2\sigma$ after considering systematic uncertainties. The product of the branching fractions of $\psip \to \gamma\etacp$ and $\etacp \to \kk\eta$ is measured to be $\BR(\psip\to\gamma \etacp)\times \BR(\etacp\to \kk\eta)=(2.39 \pm 0.32 \pm 0.34)  \times 10^{-6}$, where the first uncertainty is statistical, and the second one is systematic. The branching fraction of $\etacp \to \kk\eta$ is determined to be $\BR(\etacp \to \kk\eta) = (3.42 \pm 0.46 \pm 0.48 \pm 2.44) \times 10^{-3}$, where the third uncertainty is due to the branching fraction of $\psip \to \gamma\etacp$. Using a recent BESIII measurement of $\mathcal{B} (\etacp \to \kk\pi^{0})$, we also determine the ratio between the branching fractions of $\etacp \to \kk\eta$ and $\etacp \to \kk\pi^{0}$  to be $1.49 \pm 0.22 \pm 0.25$, which is consistent with the previous result of BaBar at a comparable precision level.

\end{abstract}

\maketitle

\section{Introduction}
Since the discovery of $J/\psi$ in 1974, the charmonium states have been viewed as excellent laboratories for studying the non-perturbative regime of Quantum Chromodynamics (QCD), which is the theory of strong interactions among quarks and gluons. Various theoretical calculations have been performed~\cite{QCD1,QCD2,QCD3,QCD4} based on QCD-inspired effective action and/or potential models. A good agreement for the mass spectrum of the charmonium states below the open-charm threshold has been achieved~\cite{QCD5,QCD6}. There are abundant measurements of the charmonium states in recent years, while our knowledge of the $S$-wave singlet charmonium state, $\eta_{c}(2S)$, is still sparse. The concise history of measurements of $\etacp$ is as follows: the resonance $\etacp$ was observed by the Belle collaboration in the decay $B^{\pm} \to K^{\pm} \eta_{c}(2S)$, $\eta_{c}(2S) \to K^{0}_{S}K^{\pm}\pi^{\mp}$~\cite{Intruduction2}, and this state was confirmed by the CLEO and BaBar collaborations in the two-photon fusion process $e^{+}e^{-} \to e^{+}e^{-}\gamma^{*}\gamma^{*},\gamma^{*}\gamma^{*} \to \eta_{c}(2S) \to K^{0}_{S}K^{\pm}\pi^{\mp}$~\cite{Intruduction3,Intruduction4}. The resonance was also observed in the double charmonium production process $e^{+}e^{-} \to J/\psi c\bar{c}$ by the BaBar~\cite{Intruduction5} and Belle~\cite{Belle} collaborations. The mass of $\eta_{c}(2S)$($M_{\etacp}= 3637.7$~MeV) lies just below that of $\psip$($M_{\psip}= 3686.1$~MeV).  Ten years after the discovery of the $\etacp$, a magnetic dipole (M1) transition $\psip \to \gamma\etacp$ was reported by the BESIII collaboration in 2012~\cite{BESIII-kskpi}, with $\etacp \to \kk \pi^{0}$ and $K_{S}^{0}K^{\pm}\pi^{\mp}$. To date, only a few $\eta_{c}(2S)$ decay modes have been observed, and the total branching fraction of these decay modes is less than 5\%~\cite{PDG}. Therefore, the search for new decay modes of $\etacp$  will provide valuable information for both experimental and theoretical studies, helping us to better understand its properties.

The search for new $\etacp$ decay modes is important as it may provide insights into unresolved charmonium puzzles. The authors of Ref.~\cite{Intruduction7} proposed that there is a “12\% rule" between the branching fractions of $J/\psi$ and $\psi(3686)$ decays:
\begin{equation}\label{12perc}
             \frac{\BR(\psip \to h)}{\BR(J/\psi \to h)} =  \frac{\BR(\psip \to e^{+}e^{-})}{\BR(J/\psi \to e^{+}e^{-})} \approx  0.12,
\end{equation}
where $h$ represents any hadronic final state.  Even though many exclusive channels obey this rule well, the decay $\psi \to \rho\pi$ has a much smaller ratio than 12\%~\cite{Intruduction8}, which gives rise to the so called “$\rho$\textendash$\pi$ puzzle”. So far, several interpretations have been proposed to solve it~\cite{rhopi1,rhopi2,rhopi3}, but none of them has been widely accepted yet. Likewise, there would be an analogous ratio between the branching fractions of $\eta_{c}(1S)$ and $\eta_{c}(2S)$, as the spin-singlet partners of $J/\psi$ and $\psip$. Ref.~\cite{Intruduction9} predicts
\begin{equation}\label{12perc-etac}
    \frac{\BR(\eta_{c}(2S) \to h)}{\BR(\eta_{c}(1S) \to h)} \approx \frac{\BR(\psip \to h)}{\BR(J/\psi \to h)} = 0.128,
\end{equation}
while Ref.~\cite{Intruduction10} predicts
\begin{equation}
    \frac{\BR(\eta_{c}(2S) \to h)}{\BR(\eta_{c}(1S) \to h)} \approx 1,
\end{equation}
if a glueball-meson mixing mechanism, which could reduce this ratio, is not taken into account. Recently, the authors of Ref.~\cite{Intruduction11} reviewed the branching fraction ratios in several decay modes, and found that seemingly most of the experimental measurements agree with neither of the two predictions. Besides, it should be noticed that the uncertainties of present experimental measurements are still very large,  preventing any definitive conclusions. The search for new decay modes and more precise measurements of $\etacp$ decays are desired to understand the hadronic decay mechanism of the charmonium states below the open-charm threshold.

Previously, evidence for $\eta_{c}(2S) \to K^{+}K^{-}\eta$ was reported by BaBar, where $\etacp$ is produced via the two-photon fusion process~\cite{Intruduction12}. The ratio between the branching fractions of $\eta_{c}(2S) \to K^{+}K^{-}\eta$ and $\eta_{c}(2S) \to K^{+}K^{-}\pi^0$ is determined to be $0.82 \pm 0.21 \pm 0.27$. With a sample of $(27.12 \pm 0.14)\times10^{8}$ $\psip$ events accumulated with the BESIII detector, we have a good chance to search for the decay $\eta_{c}(2S) \to K^{+}K^{-}\eta$, and measure its branching fraction.

\section{Detector and data samples}\label{sec:dec}
The BESIII detector~\cite{BESIIIdetector} records symmetric $e^+e^-$ collisions provided by the BEPCII storage ring~\cite{storagering} in the center-of-mass (CM) energy range from 2.0 to 4.95~GeV, with a peak luminosity of $1 \times 10^{33}~\text{cm}^{-2}\text{s}^{-1}$
achieved at $\sqrt{s} = 3.77~\text{GeV}$. BESIII has collected large data samples in this energy region~\cite{energyregion}. The cylindrical core of the BESIII detector covers 93\% of the full solid angle and consists of a helium-based multilayer drift chamber (MDC), a plastic scintillator time-of-flight system (TOF), and a CsI (Tl) electromagnetic calorimeter (EMC),
which are all enclosed in a superconducting solenoidal magnet providing a 1.0~T magnetic field.
The solenoid is supported by an octagonal flux-return yoke with resistive plate counter muon identifier modules interleaved with steel. The acceptance for charged particles and photons is 93\% over 4$\pi$ solid angle. The charged particle momentum resolution at 1 GeV/$c$ is 0.5\%,
and the specific energy loss (d$E$/d$x$) resolution is 6\% for the electrons from Bhabha scattering.
The EMC measures photon energies with a resolution of 2.5\% (5\%) at 1~GeV in the barrel (end cap) region.
The time resolution of the TOF barrel section is 68 ps,
while that of the end cap section is 110~ps. The end cap TOF system was upgraded in 2015 with multi-gap resistive plate chamber technology, providing a time resolution of 60~ps, which benefits about 85\% of the data used in this analysis~\cite{etof}.

This analysis is based on a data sample corresponding to about 2.712 billion $\psip$ events collected by the BESIII detector at the BEPCII symmetric-energy $e^{+}e^{-}$ collider. Additional data sets recorded at the center-of-mass (CM) energy of 3.650 GeV, with an integrated luminosity of 410 $\rm{pb}^{-1}$, are used to determine the non-resonant continuum background contributions. Monte Carlo (MC) simulated data samples produced with a {\sc geant4}-based~\cite{GEANT4} software package, which includes the geometric description of the BESIII detector and the detector response, are utilized to optimize event selection criteria, determine reconstruction efficiencies and estimate background contributions. The simulation models the beam-energy spread and initial state radiation (ISR) in the $e^{+}e^{-}$ annihilation with the generator {\sc kkmc}~\cite{KKMC}. The inclusive MC sample includes the production of the $\psi(3686)$ resonance, the ISR production of the $J/\psi$, and the continuum processes incorporated in {\sc kkmc}~\cite{KKMC}. The decay modes are modeled with {\sc evtgen}~\cite{EvtGen} using the known branching fractions~\cite{PDG}, and the unknown charmonium decays are modeled with {\sc lundcharm}~\cite{LUNDCHARM}. Final-state radiation (FSR) from charged final-state particles is incorporated using the {\sc photos} package~\cite{PHOTOS-package}. The exclusive decays of $\psip \to \gamma \chi_{c1,c2}$ and $\psip \to \gamma \eta_{c}(2S)$ are generated by specific models in which the angular distribution and kinematic effects have been considered. The $\etacp \to \kk\eta$ decay is generated uniformly in the phase space with the PHSP model. The $\chi_{c1} \to \kk \eta$ decays are generated as a mixture of sub-processes with intermediate states $f_{0}(980)\eta$, $f_{0}(1710)\eta$, $f_{2}'(1525)\eta$, and $K_{0}^{*}(1430)^{\pm} K^{\mp}$. For $\chi_{c2} \to \kk \eta$, the sub-processes include $f_{2}'(1525)\eta$ and $K^{+}K^{-}\eta$ in PHSP. The weights of each component are determined by fits to the invariant mass ($M_{\kk}$) spectrum of $K^{+}K^{-}$.

\section{Event Selection}
The $\etacp$ candidates are reconstructed via the decay chain $\psip \to \gamma\etacp$, $\etacp \to \kk\eta$, $\eta \to \gamma\gamma$.

Charged tracks detected in the MDC are required to be within a polar angle ($\theta$) range of $|\rm{cos\theta}|<0.93$, where $\theta$ is defined with the symmetry axis of the MDC. For charged tracks, the distance of closest approach to the interaction point must be less than 10~cm along the $z$-axis, and less than 1~cm in the transverse plane. The momentum is required to be at most 2 GeV/$c$ for each track. Charged-particle identification (PID) is based on the combined information from the specific ionization energy loss in the MDC (d$E$/d$x$) and the flight time measured by the TOF, which form the corresponding likelihood ${\cal L}(h)$ for each hadron $(h = p, K,\pi)$ hypothesis. Charged tracks are identified as kaon when the kaon hypothesis has the greatest likelihood among these hypotheses.

Photon candidates are reconstructed using isolated showers in the EMC. The deposited energy of each shower must be more than 25~MeV in the barrel region ($|\cos \theta|< 0.80$) and more than 40 MeV in the end cap region ($0.86 <|\cos \theta|< 0.92$). In this analysis, 40 MeV in the end cap region is adopted to replace the usual 50 MeV used in most of the BESIII analyses because of the low energy of photon from the M1 transition.  To exclude showers that originate from charged tracks, the angle subtended by the EMC shower and the position of the closest charged track at the EMC must be greater than 10 degrees as measured from the interaction point. To suppress electronic noise and showers unrelated to the event, the difference between the EMC time and the event start time is required to be within [0, 700]~ns.

Each candidate event is required to have one $K^{+}$ and one $K^{-}$ between three and six photon candidates.

A kinematic fit with four constraints (4C) on each $\psip \to \gamma\etacp$ candidate event is performed to suppress the backgrounds, where the total four-momentum of the final state is constrained to the initial $e^+ e^-$ four-momentum. The fit procedure loops over all photons, and the minimum fit chi-square, $\chisq_{\rm 4C}$, is used to choose the best photon candidates if the event contains more than three photons. The photon with the least energy is selected as the radiative photon, and the other two are taken as the candidates from $\eta \to \gamma\gamma$ with the $\gamma\gamma$ invariant mass being in the $\eta$ signal region, [0.51, 0.57] GeV$/c^{2}$. The backgrounds from the $\psip \to \gamma\gamma\kk\eta$ channel with four photons in the final states are suppressed by requiring $\chisq_{\rm 4C} < \chisq_{4C(4\gamma)}$.
The $\chisq_{4C(4\gamma)}$ here is the chi-square of a similar 4C kinematic fit performed on each
 $\psip \to \gamma\gamma\kk\eta$ candidate.
 The events satisfying $\chisq_{\rm 4C}<20$ are retained for further analysis. The requirements of $\chi^{2}_{\rm 4C}$ and $\eta$ signal mass region are optimized by maximizing $S/\sqrt{S+B}$, where $S$ and $B$ are the numbers of expected signal and background events determined by the MC simulation.

\begin{figure}[!ht]
  \centering
  \includegraphics[width=0.45\textwidth]{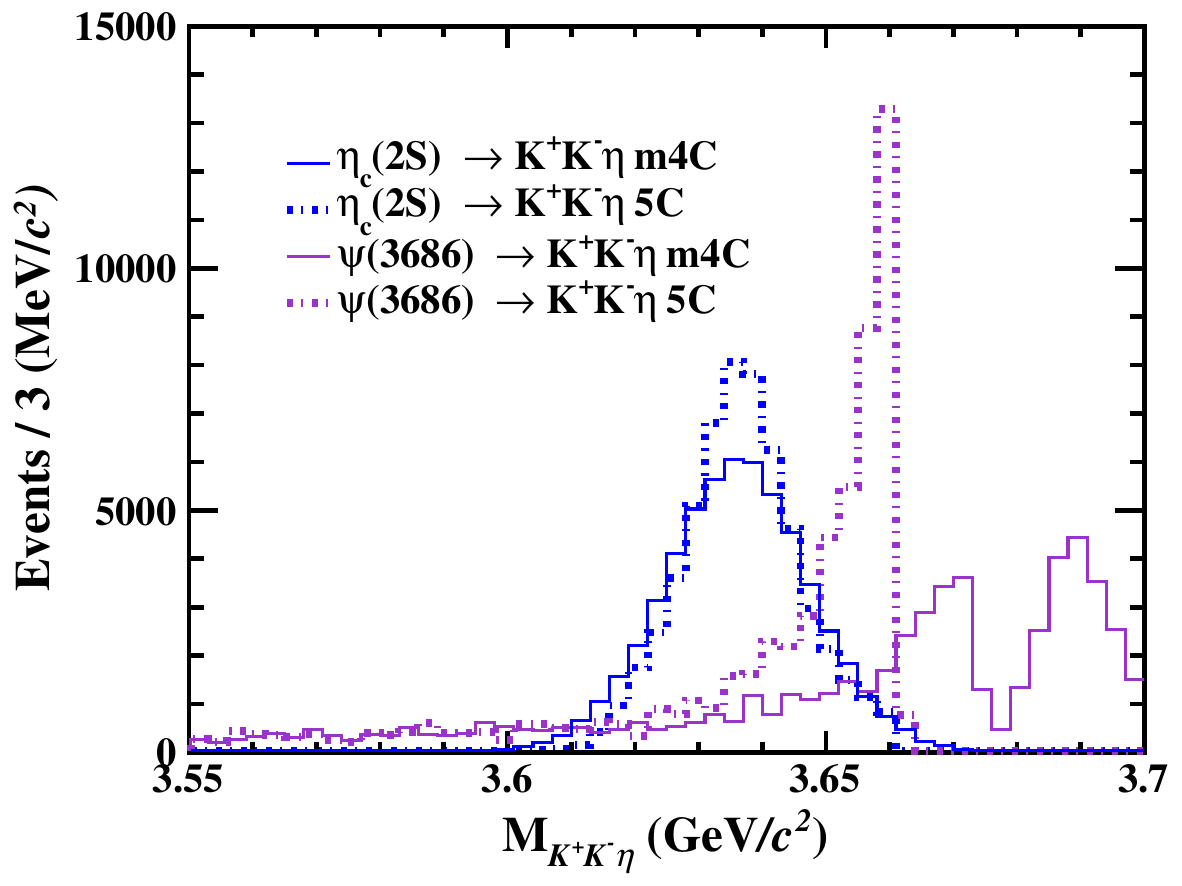}
\caption{Comparison of $M_{\kk\eta}$ between the m4C and the 5C kinematic fits of the MC samples. The blue solid and blue dash-dotted curves denote the m4C and 5C kinematic fit results of the signal channel $\psip \to \gamma \etacp \to \gamma\kk\eta$, respectively; the purple solid and purple dash-dotted curves denote the m4C and 5C kinematic fit results of $\psip \to \kk\eta$, respectively.}
\label{pic:masskketa}
\end{figure}

To improve the mass resolution, a further kinematic fit with five constraints (5C) on each $\psip \to \gamma\etacp$ candidate is performed, where the total energy-momentum of final states is constrained to the initial four-momentum and the invariant mass of the two photons is constrained to the known $\eta$  mass~\cite{PDG}. However, the background $\psip \to \kk\eta$ with a fake photon would appear as a peak close to the $\etacp$ signal in the $\kk\eta$ invariant mass ($M_{\kk\eta}$) distribution, as shown in Fig.~\ref{pic:masskketa}. This peak is suppressed in the $\etacp$ signal region by a modified 4C kinematic fit~\cite{BESIII-kskpi} (called m4C hereafter) based on the 5C kinematic fit.
The m4C configuration is the same as the 5C kinematic fit except that the energy of the radiative photon is allowed to vary in the fit.
The $M_{\kk\eta}$ distributions from the m4C are also shown in Fig.~\ref{pic:masskketa}.
In this analysis, the $M_{\kk\eta}$ distribution after applying the m4C fit is used for further study.

The decay of $\psip \to \kk\eta$ with a photon from final state radiation ($\gamma_{\rm FSR}$) is hard to be reduced because it has the same final state as the signal. The long tail of this background contaminates the signal in the $M_{\kk\eta}$ distribution. Therefore, a reliable description of its line shape is important. The contribution of this background strongly depends on the FSR ratio $R_{\rm FSR}$, which is defined as $R_{\rm FSR} = N_{\rm FSR}/N_{\rm no FSR}$, where $N_{\rm FSR}$ and $N_{\rm no FSR}$ are the numbers of events with and without the $\gamma_{\rm FSR}$~\cite{FSR-photon}. A control sample of $\psip \to \gamma\chi_{c0}\to \gamma\gamma_{\rm FSR}2(\kk)$ is selected to study the difference of $R_{\rm FSR}$ between data and MC simulation. The energy of $\gamma_{\rm FSR}$ in the control sample is between 10~MeV and 150~MeV. From the study, we find $f_{\rm FSR}= R^{\rm data}_{\rm FSR}/R^{\rm mc}_{\rm FSR} = 1.34 \pm 0.12$, where the uncertainty is statistical.
The contribution of the $\psip \to \gamma_{\rm FSR} \kk\eta$ estimated using the MC simulation is corrected to that of
data according to the measured $f_{\rm FSR}$ in the later fit.

To reduce the backgrounds associated with $\pi^{0}$, such as $\psip \to \gamma\chi_{c1}$, $\chi_{c1} \to K^{+}K^{-}\pi^{0}$, a two-dimensional veto is performed on the invariant mass distributions of $M_{\gamma_{M1}\gamma_{1}}$ and $M_{\gamma_{M1}\gamma_{2}}$. The subscript M1 represents the radiative photon, while the subscripts 1 and 2 refer to the lower and higher energy photons from the $\eta$ decay, respectively. The invariant
mass of the radiative photon from $\psip \to \gamma \chi_{c1}$ with a low-energy photon accumulates around the $\pi^0$ mass.  Therefore, events in which $\vert M_{\gamma_{M1}\gamma_{1}} - M_{\pi^{0}}\vert < 0.025$ GeV$/c^{2}$ and $\vert M_{\gamma_{M1}\gamma_{2}} - M_{\pi^{0}}\vert < 0.04$ GeV$/c^{2}$ (where $M_{\pi^{0}}=0.135$ GeV$/c^{2}$) are removed to suppress both kinds of
backgrounds. The different selection criteria here are due to the different resolutions between $M_{\gamma_{M1}\gamma_{1}}$ and $M_{\gamma_{M1}\gamma_{2}}$.
The backgrounds of $\psip \to \gamma\chi_{c1}, \chi_{c1} \to X$ (where $X$ represent $\eta f_{2}(1270), K^{+}K^{*-}, \gamma J/\psi$ and so on) are reduced by requiring $E_{\gamma_{2}}$ to be outside the range $[0.156,0.196]$ GeV.

In addition, the signal extraction suffers significantly from background contributions associated with the final state of $\gamma\kk\pi^{0}$, such as $\psip \to \gamma \eta_c$, $\eta_c \to K^+ K^- \pi^0$, and $\psip \to \gamma\chi_{c2}$, $\chi_{c2} \to K^{+}K^{-}\pi^{0}$. They cause an enhancement in $M_{3\gamma}$, the invariant mass of the three photons. We require $M_{3\gamma}>0.6 $~GeV$/c^{2}$ to suppress these backgrounds.  It should be mentioned that this requirement also removes some $\psip \to \kk \eta(\gamma_{\rm FSR})$ backgrounds, and cause a
dip in the $M_{K^+ K^- \eta}$ distribution, as shown in Fig.~\ref{pic:masskketa}.

There are some backgrounds with intermediate states of $J/\psi$ or $\phi$, such as $\psip \to \gamma\chi_{cJ}$, $\chi_{cJ} \to \gamma J/\psi$, $J/\psi \to \kk$ or $\psip \to \eta\phi$, $\phi \to \kk$.  The $\kk$ invariant mass is required to be less than $3.0$~GeV/$c^{2}$ to suppress the $J/\psi$-related backgrounds and outside the range [1.007, 1.033]~GeV$/c^{2}$ to suppress the
$\phi$-related backgrounds.

The background contribution from the continuum process (including the initial state radiation) is estimated using the data set taken at the CM energy of 3.650~GeV with an integrated luminosity of 401.00 $\rm {pb}^{-1}$. The mass spectrum is scaled due to the difference in the CM energies, and the number of events is normalized based on the corresponding integrated luminosities~\cite{psip3year} and cross-sections.
The normalized number of continuum events is about two in the region of $M_{K^+K^-\eta}\in [3.60,3.65]$ ${\rm GeV}/c^2$.

\section{Signal extraction and branching fraction}
The signal yields are obtained by a fit to the $M_{\kk\eta}$ spectrum with an unbinned maximum likelihood method. The fit is performed in the mass range of $[3.45,3.70]$~GeV/$c^{2}$ to cover the $\chi_{c1,c2}$ signals. The line shape of $\etacp$ is described by~\cite{BAM-00500, kkpi0}
\begin{equation}
    (E_{\gamma}^{3} \cdot BW(M) \cdot f_{d}(E_{\gamma}) \cdot \epsilon(M)) \otimes DG_{\rm{all-resolution}},%
\end{equation}
where $M$ denotes $M_{\kk\eta}$, $E_{\gamma}$ is the energy of the transition photon in the
$\psip$
rest frame, which is taken as
\begin{equation}
E_{\gamma} = \frac{m^{2}_{\psip}-M^{2}}{2m_{\psip}}.
\end{equation}
The Breit-Wigner function for $\eta_{c}(2S)$ is taken as
\begin{equation}
BW(M) =\frac{1}{|M^2-m^2_{\eta_{c}(2S)}+i m_{\eta_{c}(2S)} \Gamma_{\eta_{c}(2S)}|^{2}},
\end{equation}
with the $\etacp$ mass and width fixed to the world average values~\cite{PDG}. The $f_d(E_\gamma)$ is a damping function proposed by the KEDR experiment~\cite{KEDR}, which is written as
\begin{equation}
f_{d}(E_{\gamma}) = \frac{E_{0}^{2}}{E_{\gamma}E_{0} + (E_{\gamma}-E_{0})^{2}},
\end{equation}
to suppress the diverging tail caused by the factor $E_{\gamma}^{3}$. The nominal energy of the transition photon, $E_0$, is calculated with
\begin{equation}
E_{0} = \frac{m^{2}_{\psip}-m^{2}_{\eta_{c}(2S)}}{2m_{\psip}}.
\end{equation}
The efficiency $\epsilon(M)$ is a function of $M_{\kk\eta}$ determined by MC simulation, which is fitted with a 5th order Chebyshev function. $DG_{\rm{all-resolution}}$ is a Gaussian-like function describing the resolution. It accounts for both the detector resolution, represented by a double Gaussian function with parameters determined by MC simulation, and the discrepancy between data and MC simulation, represented by a single Gaussian function with parameters obtained from control samples of $\chi_{c1,c2} \to K^+ K^- \eta$.
To incorporate both effects, a new function $DG_{\rm{all-resolution}}$ is constructed by convolving the double Gaussian function with a single Gaussian function. All parameters of $DG_{\rm{all-resolution}}$ are fixed in the fit.

The $\chi_{c1}$ and $\chi_{c2}$ signals are described by the simulated MC shapes convolved with a single Gaussian function to take into account the resolution differences in $M_{\kk\eta}$ between data and MC simulation.

To determine the contribution of the continuum background, we fit the shape of the shifted mass spectrum with a second-order Chebyshev function. In the nominal fit, the parameters and signal yields of the continuum background are fixed. For $\psip \to (\gamma_{\rm FSR})\kk\eta$, signal shape is derived from the signal MC sample with corrected $f_{\rm FSR}$. All the other smooth backgrounds are described with an ARGUS function~\cite{Argus}, with the threshold parameter fixed at 3.700~GeV, while the other parameters are left free in the fit.

\begin{figure*}[htbp]
  \centering
  \subfigure[]{
  \label{Fig.sub.1}
  \includegraphics[width=0.49\textwidth]{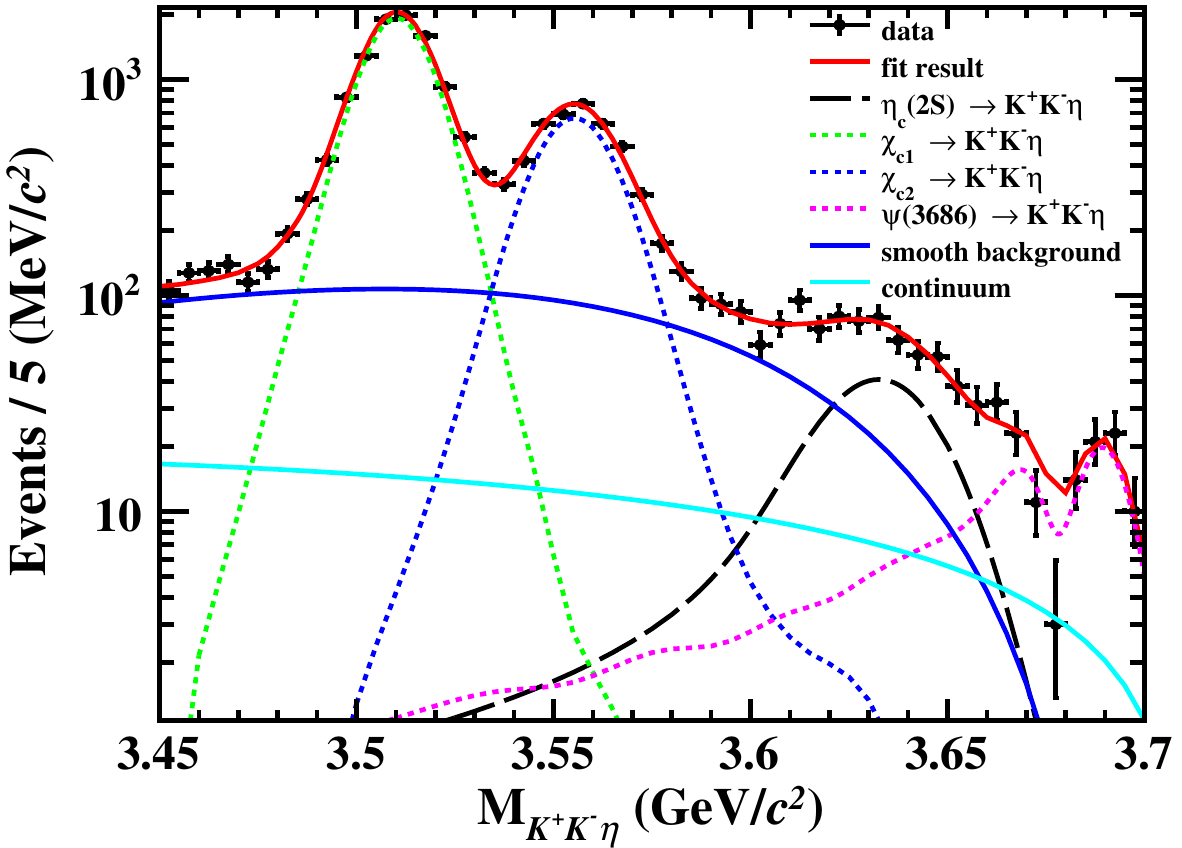}}
  \subfigure[]{
  \label{Fig.sub.2}
  \includegraphics[width=0.49\textwidth]{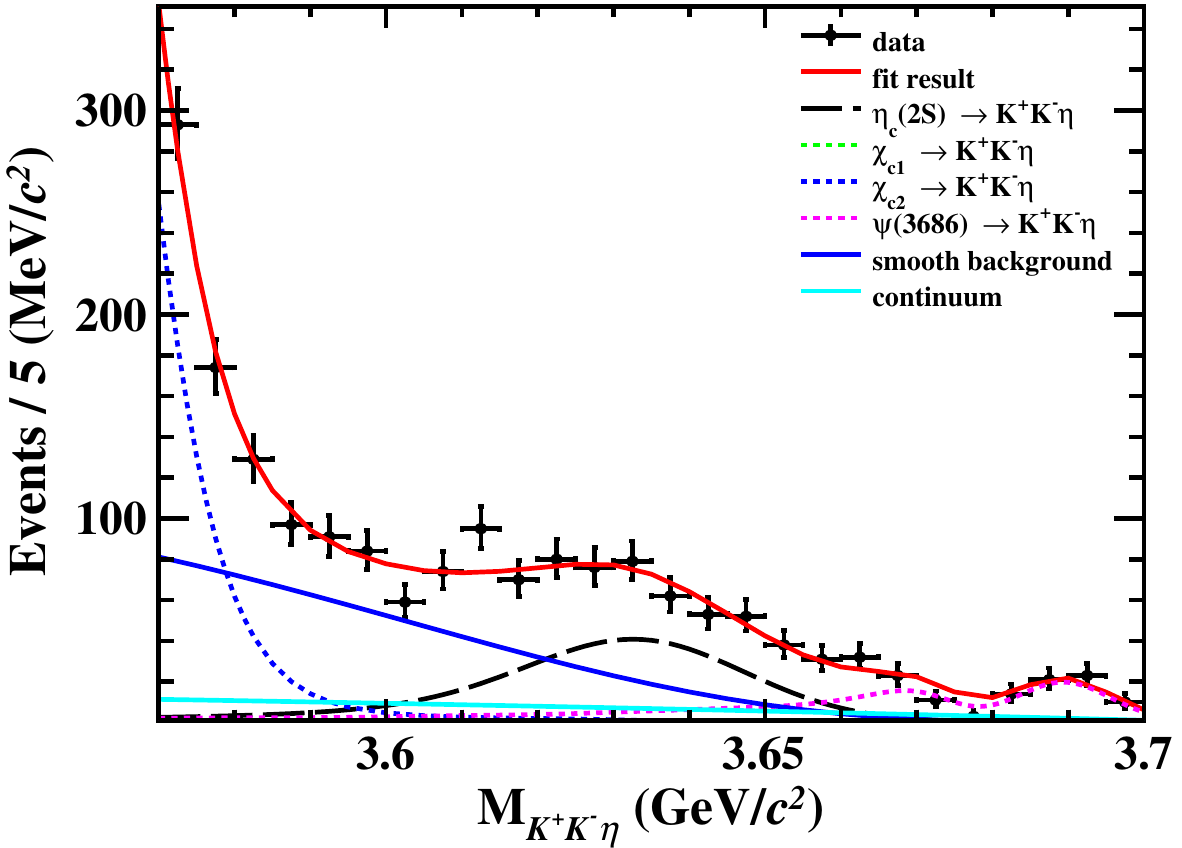}}
\caption{Results of the fit to the invariant mass distribution of $\kk\eta$, illustrated in the whole fit range (left) and the range only containing the $\etacp$ signal (right). Dots with error bars are data. The red, blue, and cyan solid curves are the total fit, smooth background, and continuum background, respectively. The green dotted, blue dotted, and black dash-dotted curves denote decay modes of $\chi_{c1}, \chi_{c2}$, and $\etacp$, respectively. The pink dotted curve is the $\psip \to (\gamma_{\rm FSR})\kk\eta$ contribution.}
\label{pic:fitresult}
\end{figure*}

In the fit, it is assumed that there is no interference between the signal and continuum amplitudes. The fit results are shown in Fig.~\ref{pic:fitresult}. The signal yields from the fit are summarized in Table~\ref{fitrsult}. The $\chisq/ndf$ value of the fit is $79.6/44$, where $ndf$ is the number of degrees of freedom. This relatively large $\chisq/ndf$ is primarily due to the large statistics of the $\chi_{c1,c2}$ data samples.  The significance of the $\etacp$ signal, calculated by the difference of the likelihoods and the $ndf$ with and without the $\etacp$ signal
component in the fit~\cite{likelihood}, is determined to be 6.2$\sigma$ after considering systematic uncertainties.

An input-output check using 300 toy MC samples shows that the fit scheme is stable and do not induce bias.

The branching fraction of the $X \to \kk\eta$ decay is determined by
\begin{equation}
    \BR(X \to \kk\eta) = \frac{N_{X}}{N_{\psip} \cdot \epsilon \cdot \BR_{0}},
\end{equation}
where $X$ represents $\chi_{c1}$, $\chi_{c2}$ or $\etacp$, $N_{X}$ is the number of observed signal events, $N_{\psip}$ is the total number of $\psip$ events, $\epsilon$ is the detection efficiency after corrections as described in Sec.~\ref{uncertainty}, $\BR_{0}$ is the product of the branching fractions $\psip \to \gamma X$ and $\eta \to \gamma \gamma$~\cite{PDG}. Table~\ref{fitrsult} lists the fit results, branching fractions, and other relevant values.

\begin{table*}[htbp]
	\centering
	\caption{The signal yields ($N_X$), the detection efficiencies ($\epsilon$), the obtained branching fractions $(\mathcal B)$, as well as comparison to the PDG values.  Only statistical uncertainties are presented, except those quoted from PDG.}
 	\begin{tabular}{c@{\hspace{1cm}}c@{\hspace{1cm}}c@{\hspace{1cm}}c}
	\hline\hline	  Channel &	$\eta_{c}(2S) \to K^{+}K^{-}\eta$ & $\chi_{c1} \to K^{+}K^{-}\eta$ & $\chi_{c2} \to K^{+}K^{-}\eta$	~\\	\hline
	$N_X$		&	$362 \pm 49$  &	$8690 \pm 110$ &	$3693\pm 85$ 	~\\
	$\epsilon$ (\%)		&	$14.2$	 &	$36.6$	&	$20.7$		~\\	
	$\mathcal B$ & $(3.25 \pm 0.44)\times 10^{-3}$	& $(2.38 \pm 0.03)\times 10^{-4}$		& $(1.75 \pm 0.04) \times 10^{-4}$	~\\
	${\mathcal B}^{\rm PDG}$ ~\cite{PDG}	& $(5 \pm 4) \times 10^{-3} $ 	& 	$(3.2 \pm 1.0) \times 10^{-4}$	& 	$<3.2\times 10^{-4}$ ~\\	 \hline
	\end{tabular}
	\label{fitrsult}
\end{table*}

\section{SYSTEMATIC UNCERTAINTY}\label{uncertainty}
The sources of systematic uncertainties considered in the branching fraction measurement include the total number of $\psip$ events, tracking, PID, photon detection, branching fractions of the intermediate decays, kinematic fit, misidentification of photons, veto criteria with photon, veto criteria on $M_{\kk}$, and the fit.

The total number of $\psip$ events is determined with inclusive hadronic $\psip$ decays. With the same method in Ref.~\cite{psip3year}, in which the $\psip$ data sample collected in 2009 and 2012 is analysed, the total number of $\psip$ events collected in 2009, 2012, and 2021 is determined to be $(27.12\pm 0.14) \times 10^{8}$ with an uncertainty of 0.5\%.

Using the control sample of $e^{+}e^{-} \to \pi^{+}\pi^{-}K^{+}K^{-}$, the uncertainties of the tracking and PID are determined to be  1.0\% for each kaon~\cite{kaon-track}.

The uncertainty due to photon reconstruction is determined to be 1.0\% per photon using the control sample of  $J/\psi \to \pi^{+}\pi^{-}\pi^{0}$~\cite{photon}.

The systematic uncertainties due to the branching fractions of $\psip \to \gamma\eta_{c}(2S)$ and $\eta \to \gamma\gamma$ are quoted as 71.4\% and 0.5\%~\cite{PDG}, respectively.

To study the uncertainty associated with the 4C kinematic fit, we correct the track helix parameters in the MC simulation. The helix correction parameters of charged kaon are obtained with the control sample of $\psip \to \gamma \chi_{c0} \to \gamma2(K^{+}K^{-})$. The resulting efficiency difference before and after the correction, 2.1\%, is taken as the systematic uncertainty.

The systematic uncertainty associated with the misidentification of a photon is estimated by utilizing the control sample of $\psip \to \eta J/\psi$. The efficiencies of reconstructing $\eta$ obtained from MC simulation are corrected according to the data-MC difference. The correction factor is determined to be $f = 0.953\pm 0.005$. The residual uncertainty of 0.5\% is taken as the systematic uncertainty after efficiency correction.

To estimate the systematic uncertainty of the $\eta$ signal interval selection and veto criteria associated with photons, we smear the energy of photons based on the difference in the resolutions between data and MC simulation. The relevant systematic uncertainty is estimated to be 2.4\% based on the difference of the efficiencies before and after the smearing. The uncertainty from the $M_{K^{+}K^{-}}$ criteria, that require the invariant mass of $\kk$($M_{\kk}$) to be outside the range of [1.007, 1.033]~GeV$/c^{2}$ and less than 3.0~GeV$/c^{2}$ to veto $\phi$ and $J/\psi$, is estimated by comparing the final results with and without applying the criteria. We find a difference of 0.5\% and take it as the systematic uncertainty due to the $M_{K^{+}K^{-}}$ criteria.

The systematic uncertainties associated with the signal and background shapes in the fit to the $M_{\kk\eta}$ distribution are estimated as follows. The systematic uncertainties due to the line shape of the smooth backgrounds are estimated with an alternative line shape, described by a mixing-component shape from $\eta$ sideband and flat background with $\eta$ from the inclusive MC sample. The change of the fitted signal yield, 5.0\%, is assigned as the uncertainty. The systematic uncertainties associated with the resonance parameters of $\etacp$ are estimated to be 1.2\% and 11.9\%, respectively, by changing the $\etacp$ mass and width by $\pm 1\sigma$ according to the world average values~\cite{PDG}. The systematic uncertainty of the fit range is determined to be negligible  using the Barlow test~\cite{Barlow}. The systematic uncertainty from the damping factor is estimated to be 0.5\%, by changing the invariant mass of $\etacp$, which is a parameter in the damping factor function, by $\pm 1\sigma$. The systematic uncertainty due to the background $\psip \to \kk\eta$ with $\gamma_{\rm FSR}$ is estimated by varying $f_{\rm FSR}$ by $\pm 1\sigma$. The resulting difference to the nominal one, 1.4\%, is assigned as the corresponding uncertainty. The systematic uncertainty due to the continuum process is found to be negligible by changing the number of continuum background events within $\pm 1\sigma$.

An alternative Chebyshev function, changed from 5th-order to 6th-order to fit the efficiency curve, is chosen to estimate the uncertainty from the efficiency curve. The difference between the two functions, 0.3\%, is taken as the systematic uncertainty.

Among all sources of systematic uncertainties, the largest one comes from the quoted branching fraction of $\psip \to \gamma\etacp$. Since it is not possible to reduce that in this analysis, we treat it separately. All the other sources of systematic uncertainties are assumed to be independent of each other and are combined in quadrature to obtain the overall systematic uncertainty as listed in Table~\ref{total sys err}.

\begin{table}[!htbp]
\caption{\label{total sys err} The relative systematic uncertainties (in \%) on the branching fraction measurement.}
\begin{tabular}{cc}
    \hline \hline
    Source&        Uncertainty   \\   \hline
    $N_{\psip}$ & 0.5   \\
  Continuum contribution  & 1.0 \\
     Tracking        &2.0         \\      PID & 2.0 \\
    Photon reconstruction                   &3.0       \\
    $\BR(\eta \to \gamma\gamma)$ & 0.5        \\
    4C kinematic fit                           &2.1       \\
    Photon misidentification & 0.5 \\
    Veto criteria with photon                     &  2.4    \\
     $M_{\kk}$ requirement  & 0.5\\
     Background shape & 5.0 \\
     Mass of $\etacp$  & 1.2\\
     Width of $\etacp$  & 11.9\\
     Damping factor & 0.5 \\
     Ratio of FSR  & 1.4 \\
     Efficiency curve  & 0.3 \\
    \hline
    Total                                         &14.1      \\  \hline
    $\BR(\psip \to \gamma\etacp)$  &  71.4 \\
    \hline \hline
\end{tabular}
\end{table}

\section{SUMMARY and discussions}
Using the $(27.12\pm0.14)\times10^{8}$ $\psip$ events collected by the BESIII detector, the $\eta_{c}(2S) \to K^{+}K^{-}\eta$ decay is observed for the first time.  The statistical significance of the signal is $6.2\sigma$ after considering systematic uncertainties.
The product of the branching fractions is determined to be $\BR(\psip\to \gamma\etacp)\times \BR (\etacp \to \kk\eta)=(2.39 \pm 0.32 \pm 0.34) \times 10^{-6}$, where the first uncertainty is statistical and the second systematic. The branching fraction of $\eta_{c}(2S) \to K^{+}K^{-}\eta$ is calculated to be $(3.42 \pm 0.46 \pm 0.48 \pm 2.44) \times 10^{-3}$, with an additional third uncertainty coming from the quoted $\BR(\psip \to \gamma \etacp)$. The branching fraction of $\eta_{c}(2S) \to K\bar{K}\eta$ is calculated to be $(6.84 \pm 0.92 \pm 0.96 \pm 4.88) \times 10^{-3}$ based on the isospin symmetry.

Furthermore, with the recent BESIII measurement of ${\cal B} (\eta_{c}(2S) \to K^+ K^- \pi^0$)~\cite{kkpi0}, the ratio between the $\BR(\eta_{c}(2S) \to K^{+}K^{-}\eta)$ and $\BR(\eta_{c}(2S) \to K^{+}K^{-}\pi^0)$ is determined to be $1.49 \pm 0.22 \pm 0.25$. Our result is consistent with the BaBar result $0.82 \pm 0.21 \pm 0.27$~\cite{Intruduction12}.
\begin{figure}[htbp]
  \centering
  \includegraphics[width=0.45\textwidth]{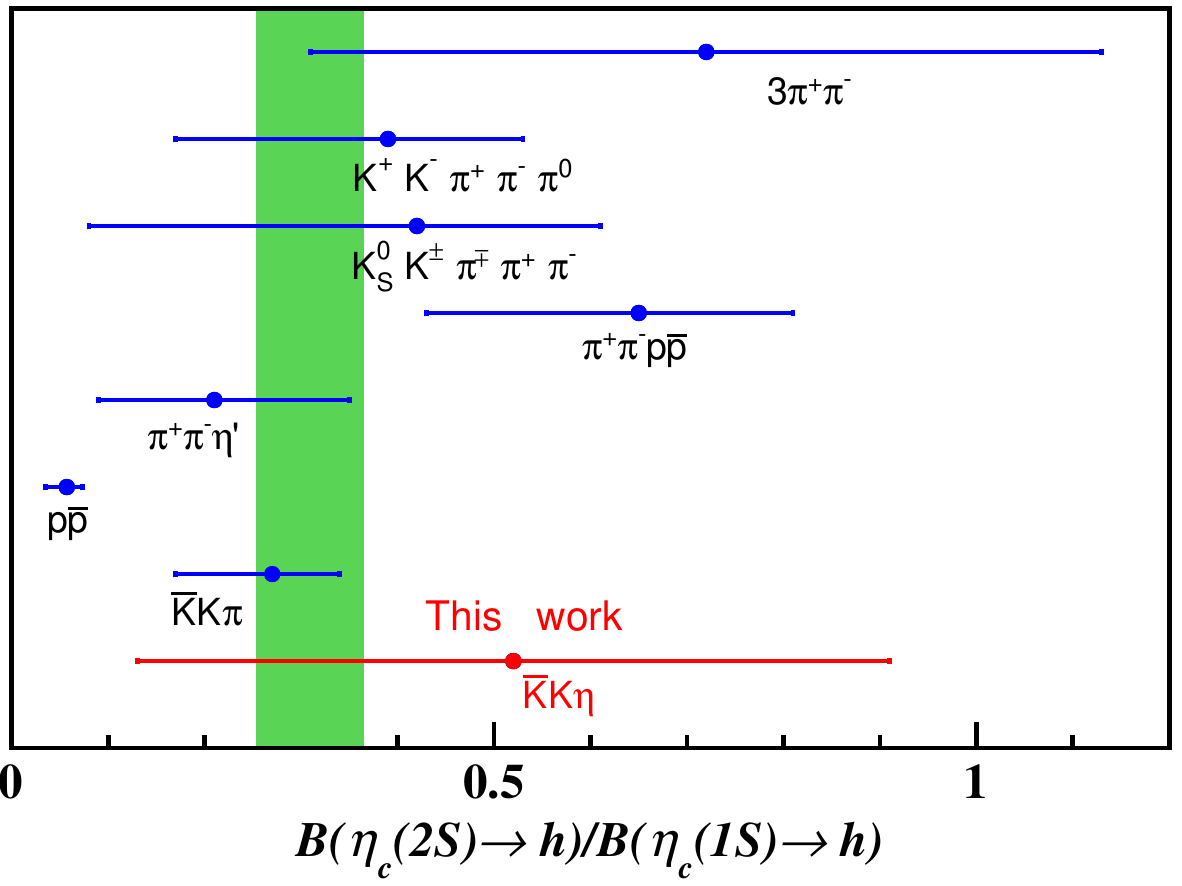}
\caption{Comparison of ${\cal B}(\etacp \to h)$/${\cal B}(\eta_{c}(1S) \to h)$. Except for the result of the channel $K\bar{K}\eta$ obtained in this work, the others are quoted from Ref.~\cite{Intruduction11}. The green band represents the result of fitting the ratios for the decay modes excluding the $p\bar{p}$
channel.}
\label{pic:ratio}
\end{figure}
With the branching fraction of $\eta_{c}(1S) \to \kk\eta$, the ratio of the branching fractions of $\eta_{c}(1S)$ and $\etacp$ decaying into $\kk\eta$ is calculated to be ${\cal B} (\etacp \to K\bar{K}\eta)$/${\cal B}(\eta_{c}(1S) \to K\bar{K}\eta)= 0.52 \pm 0.39$. Comparing this to the ratios from other hadronic decay modes of $\etacp$/$\eta_{c}(1S)$~\cite{Intruduction11} shown in  Fig.~\ref{pic:ratio}, it is observed that the averaged value is around 0.3, which does not align with the predictions in either Ref.~\cite{Intruduction9} or in Ref.~\cite{Intruduction10}. The observed discrepancy reflects our limited knowledge of the decay mechanisms of the spin singlet charmonium states. More searches on new decay modes and more precise measurements of the $\etacp$ decays are required to clarify this puzzle.

\section{Acknowledgement}
The BESIII Collaboration thanks the staff of BEPCII and the IHEP computing center for their strong support. This work is supported in part by National Key R\&D Program of China under Contracts Nos. 2020YFA0406300, 2020YFA0406400, 2023YFA1606000; National Natural Science Foundation of China (NSFC) under Contracts Nos. 12275058, 11635010, 11735014, 11835012, 11935015, 11935016, 11935018, 11961141012, 12025502, 12035009, 12035013, 12061131003, 12192260, 12192261, 12192262, 12192263, 12192264, 12192265, 12221005, 12225509, 12235017; the Chinese Academy of Sciences (CAS) Large-Scale Scientific Facility Program; the CAS Center for Excellence in Particle Physics (CCEPP); Joint Large-Scale Scientific Facility Funds of the NSFC and CAS under Contract No. U1832207; CAS Key Research Program of Frontier Sciences under Contracts Nos. QYZDJ-SSW-SLH003, QYZDJ-SSW-SLH040; 100 Talents Program of CAS; The Institute of Nuclear and Particle Physics (INPAC) and Shanghai Key Laboratory for Particle Physics and Cosmology; European Union's Horizon 2020 research and innovation programme under Marie Sklodowska-Curie grant agreement under Contract No. 894790; German Research Foundation DFG under Contracts Nos. 455635585, Collaborative Research Center CRC 1044, FOR5327, GRK 2149; Istituto Nazionale di Fisica Nucleare, Italy; Ministry of Development of Turkey under Contract No. DPT2006K-120470; National Research Foundation of Korea under Contract No. NRF-2022R1A2C1092335; National Science and Technology fund of Mongolia; National Science Research and Innovation Fund (NSRF) via the Program Management Unit for Human Resources \& Institutional Development, Research and Innovation of Thailand under Contract No. B16F640076; Polish National Science Centre under Contract No. 2019/35/O/ST2/02907; The Swedish Research Council; U. S. Department of Energy under Contract No. DE-FG02-05ER41374.


\end{document}